# Botnet-based Distributed Denial of Service (DDoS) Attacks on Web Servers: Classification and Art

[1]Esraa Alomari,
[2]Selvakumar Manickam
[1,2]National Advanced IPv6 Centre (NAV6), Universiti Sains Malaysia, Malaysia

[3,4]B. B. Gupta
[3]University of New Brunswick, Canada
[4]RSCOE, University of Pune, India

[5]Shankar Karuppayah,
[6]Rafeef Alfaris
[5,6]National Advanced IPv6 Centre (NAV6), Universiti Sains Malaysia, Malaysia

## ABSTRACT
Botnets are prevailing mechanisms for the facilitation of the distributed denial of service (DDoS) attacks on computer networks or applications. Currently, Botnet-based DDoS attacks on the application layer are latest and most problematic trends in network security threats. Botnet-based DDoS attacks on the application layer limits resources, curtails revenue, and yields customer dissatisfaction, among others. DDoS attacks are among the most difficult problems to resolve online, especially, when the target is the Web server. In this paper, we present a comprehensive study to show the danger of Botnet-based DDoS attacks on application layer, especially on the Web server and the increased incidents of such attacks that has evidently increased recently. Botnet-based DDoS attacks incidents and revenue losses of famous companies and government websites are also described. This provides better understanding of the problem, current solution space, and future research scope to defend against such attacks efficiently.

## General Terms
Information security, Computer network.

## Keywords
Information security, Botnet, DDoS attacks, IRC, Web server.

## 1. INTRODUCTION
The rapid development of the Internet over the past decade appeared to have facilitated an increase in the incidents of online attacks [1]. One such powerful and harmful attack is the denial of service (DoS) attack. A DoS attack significantly threatens the network, especially if such an attack is distributed. A distributed DoS (DDoS) attack is launched by a mechanism called Botnet through a network of controlled computers. A software program controls the computers and for specific purposes, known as "bots." Bots are small scripts that have been designed to perform specific, automated functions. Bots are utilized by agents for Web indexing or "spidering," as well as to collect online product prices or to performing such duties as chatting. However, bots are negatively associated with "remote access Trojan Horses" (e.g., Zeus bot) and zombie computers that are created for less favorable purposes [2]. Bots in large quantities provide the power of a computer to create prime tools for such activities as the widespread delivery of SPAM email, click-fraud, spyware installation, virus and worm dissemination, and DDoS attacks (e.g., black energy bot) [3]. DDoS attacks usually take advantage of the weaknesses of a network layer, particularly, SYN, UDP, and Internet control message protocol (ICMP) flooding. Such attacks encroach the network bandwidth and resources of the victim, thus facilitating the denial of legitimate access.

A DDoS attack is exemplified by the direct attempt of attackers to prevent legitimate users from using a specific service [4]. A recent, sophisticated, and popular method of DDoS attack involves application level flooding, especially in the Web server. Such attacks employ various flooding methodologies (e.g., HTTP-GET flood, etc). Figure 1 shows the types of DDoS attacks on the application layer in 2010 and 2011, as reported by the Arbor Inc. [5]. From the figure, we can see that HTTP attacks rank first in terms of number of incidents. HTTPs registered the highest incidence of DDoS attacks in 2010, reaching up to 100 Gbps in 2011. This increase accounts for a 700% rise in incidents, as reported by the Cloud Flare Company [6], where the HTTP attacks comprise approximately 80% in 2010, a value that significantly increased to approximately 88% in 2011. The number of daily target Web sites evidently increased, with government websites becoming a common target [5].

The remainder of the paper is organized as follows. Section 2 contains overview of the Botnet based DDoS attacks. Section 3 presents Botnet based DDoS Attack Architecture. Botnet based DDoS attack tools are described in section 4. Section 5 describes classification of Botnet based DDoS attacks in details. Section 6 contains various Botnet based DDoS attack incidents. Finally, Section 7 concludes the paper and presents further research scope.

## 2. BOTNETS BASED DDOS ATTACKS
This section provides a background on Botnets and how they facilitate DDoS attacks that hamper the Web server. Botnets compromise a network of machines with programs (usually referred to as a bot, zombie, or drone) and implement under a command and control (C&C) management infrastructure. The management of Botnets typically affects a series of systems through numerous tools and through the installation of a bot that can remotely control the victim using Internet relay chat (IRC) [7]. Present Botnets are most frequently used to spread DDoS attacks on the Web. Moreover, the attackers can change their communication approach during the creation of the bots. Majority of bots varied its potentials to participate in such attacks. The most typical and commonly implemented Botnet attack on application layer is the HTTP/S flooding attack, which launches bots created by the HTTP server. Such bots are thus called "Web-based" bots [8].

The goal of a Botnet based DDoS attack is to entail damage at the victim side. In general, the ulterior motive behind this attack is personal which means block the available resources or degrade the performance of the service which is required by the target machine. Therefore, DDoS attack is committed for the revenge purpose. Another aim to perform these attacks can be to gain popularity in the hacker community. In addition to this, these attacks can also perform for the material gain, which means to break the confidentiality and use data for their use.

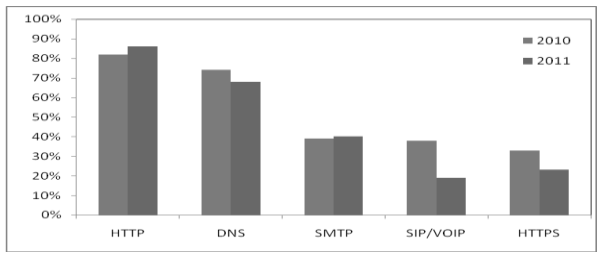

**Fig. 1: Types of DDoS attacks on the application layer**

## 3. BOTNET BASWD DDOS ATTACK ARCHITECTURE

Botnet based DDoS attack networks fall under three categories, namely, the agent-handler, IRC-based, and Web-based models.

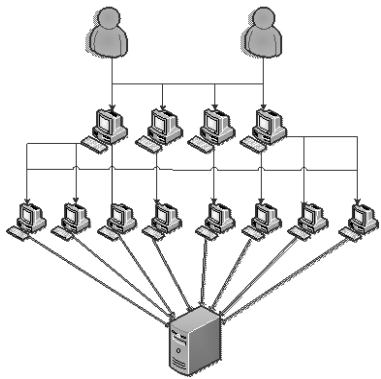

**Fig. 2: Agent–Handler Model**

### 3.1 Agent-Handler Model

The agent-handler model of a DDoS attack comprises clients, handlers, and agents as shown in Figure 2. The client is one with whom the attacker communicates in the DDoS attack system. The handlers are software packages located throughout the Internet. The client uses these packages to communicate with the agents. The agent software thrives in compromised systems, eventually conducting the attack at the appropriate time.

The attacker communicates with any of the handlers to identify operational agents and to determine when to attack or to upgrade agents. Owners and users of agent systems are typically unaware that their system has been compromised and is under a DDoS attack. Depending on the configuration of the DDoS attack network, agents can be instructed to communicate with one handler or with multiple handlers. Attackers often attempt to install the handler software on a compromised router or network server. The target typically handles large volumes of traffic, making message identification difficult between the client and the handler and between the handler and the agents. The terms "handler" and "agents" are sometimes replaced with "master" and "demons," respectively, in descriptions of DDoS tools [9].

### 3.2 Internet Relay Chat (IRC) Model

The architectures of the IRC-based DDoS attack as shown in Figure 3 and of the agent–handler model are almost similar. However, instead of employing a handler program that is installed on a network server, the client is connected to the agents through an IRC communication channel. An IRC channel benefits an attacker with the use of "legitimate" IRC ports to send commands to agents. The use of legitimate ports hinders the tracking DDoS command packets. Additionally, IRC servers tend to have large volumes of traffic, enabling an attacker to conceal its presence easily. The attacker does not necessarily maintain a list of the agents because it can immediately enter the IRC server and view all available agents [10]. The agent software in the IRC network sends and receives messages through the IRC channel and informs the attacker when an agent becomes operational.

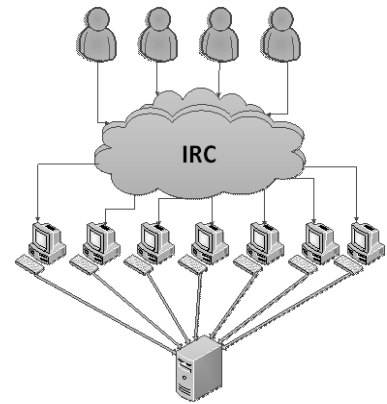

**Fig. 3: Internet Relay Chat (IRC) Model**

### 3.3 Web-based Model

Although the most preferred method for Botnet command and control (C&C) is the IRC-based model, Web-based reporting and command has emerged over the past few years. A number of bots in the Web-based model simply report statistics to a Web site, whereas others are intended to be fully configured and controlled through complex PHP scripts and encrypted communications over the 80/443 port and the HTTP/HTTPS protocol. The following are the advantages of Web-based controls over IRC [11]:

- Ease of set-up and website configuration;

- Improved reporting and command functions;

- Less bandwidth requirement and the acceptance of large Botnets for the distributed load;

- Concealment of traffic and hindrance of filtering through the use of port 80/443;

- Resistance to Botnet hijacking via chat-room hijacking; and

- Ease of use and of acquisition.

## 4. BOTNETS BASED DDOS ATTACK TOOLS

Various DDoS attack tools are known and architectures are very similar that some tools actually originate from minor modifications of other tools [12]. In this section, the functionality of a number of these tools is discussed. The Botnet based DDoS attack tools are classified as agent-based, IRC-based, or Web-based DDoS attack tools.

### 4.1 Agent-based DDoS Attack Tools

Agent-based DDoS attack tools are based on the agent–handler DDoS attack model comprising handlers, agents, and victims, as described in Section 3.1. Examples of agent-based DDoS tools are Trinoo, Tribe Flood Network (TFN), TFN2K, Stacheldraht, Mstream, and Shaft [13]. Among the abovementioned agent-based DDoS tools, Trinoo [14] is the most popular and the most widely used for its capability for bandwidth depletion and for launching UDP flood attacks against one or numerous Internet protocol (IP) addresses. Shaft [15], on the other hand, is similar to Trinoo in that it can launch packet flooding attacks. Shaft can also control the duration of the attack, as well as the size of the flooding packets.

TFN [16] is another DDoS attack tool that can conduct bandwidth and resource depletion attacks. TFN can perform Smurf, UDP flooding, TCP SYN flooding, ICMP echo request flooding, and ICMP directed broadcast. TFN2K [15], as a derivative of TFN, can perform Smurf, SYN, UDP, and ICMP flood attacks. TFN2K has the special capability of adding encrypted messages between attack components. Stacheldraht [17] is a product of previous TFN attempts. Stacheldraht strengthens a number of TFN's weak points and is capable of implementing Smurf, SYN flood, UDP flood, and ICMP flood attacks. On the other hand, Mstream [18] is a simple point-to-point TCP ACK flooding tool that can overwhelm fast-routing routine tables in some switches.

### 4.2 IRC-based DDoS Attack Tools

IRC-based DDoS attack tools were developed after the emergence of agent–handler attack tools. More sophisticated IRC-based tools have been developed, and these tools include the important features of several agent-handler attack tools. The Trinity is one of the best-known IRC-based DDoS tools on top of UDP, TCP SYN, TCP ACK, and TCP NUL packet floods. The Trinity v3 [19] introduces TCP random flag packet floods, TCP fragment floods, TCP established floods, and TCP RST packet floods. Along with the development of the Trinity came the myServer [15], that rely on external programs to conduct DoS and plague to simulate TCP ACK and TCP SYN flooding. Knight [20] is another light-weight and powerful IRC-based DDoS attack tool that can perform UDP flood attacks and SYN attacks. Knight can be considered an urgent pointer flooder [9]. An IRC-based DDoS tool based on Knight is Kaiten [20], which conducts UDP, TCP flood attacks, SYN, and PUSH+ACH attacks.

### 4.3 Web-based DDoS Attack Tools

Web-based DDoS attack tools were recently developed with the purpose of attacking the application layer, especially the Web server. IRC-based DDoS attack tools with the HTTP/S flooding function are used to attack a Web server, thus proving that attackers are increasingly adopting various tools to introduce DDoS attacks [5].

Unlike currently popular attack tools that can launch DDoS attacks, most organizations are unaware of the broad development over the last few years and are vulnerable to attackers, according to the Arbor Networks. Commercial services, along with downloadable tools, can launch attacks for a fee [5]. Therefore, we discuss the bot tools that launch DDoS attacks on the application layer. Approximately 20,000 infected computers with multiple targets can destroy over 90% of Internet sites [21]. A DDoS attack on the application layer is highly comparable to calling someone in the world from one Website, while the Web site indicates being out of service or displays "the page cannot found." Therefore, the server hosting the site cannot process all requests on the same site, in contrast to the compromising handler that injects the site with bots controlled by attackers. The attacker consequently demonstrates the use of different tools to execute a successful attack. In the following sections, three Web-based DDoS attack tools are described.

*4.3.1 BlackEnergy*
BlackEnergy [60] is a Web-based DDoS bot used by unidentified Russian hackers. BlackEnergy easily controls Web-based bots through minimal syntax and structure, resulting in the launch of various attacks. One or more Russian hackers had apparently developed this tool. Meanwhile, most BlackEnergy C&C systems are seen in Malaysia and in Russia, with Russian sites being the primary targets. One of the main features that BlackEnergy bot promote in forums is the capability to target more than one IP address per host name. This tool continues to be widely used to deny services from commercial Web sites.

*4.3.2 Low-Orbit Ion Cannon (LOIC)*
The LOIC is a Botnet-based DDoS attack tool that releases flooding in the server. This flooding apparently results from the large volume of HTTP traffic. However, this tool has been used recently by an anonymous group to facilitate malicious traffic through the Zeus Botnet, which is an advanced malware program that cannot be easily removed. The hacker group administered the largest attack in 2012 against famous Web sites, such as the Department of Justice (DOJ) and the Federal Bureau of investigation (FBI) [22].

*4.3.3 Aldi Botnet*
Aldi is a newer inexpensive DDoS bot that is growing in popularity. Recent data [59] suggests that there are at least 50 distinct Aldi bot binaries that have been seen in the wild with 44 unique Command & Control (C&C) points. As per Arbor company which monitors real time Internet traffic, this bot is active in Russia, Ukraine, US and Germany.

## 5. CLASSIFICATION OF BOTNETS BASED DDOS ATTACKS

The wide variety of DDoS attacks indicates the various conducted taxonomies of such attacks [9, 23-27]. New kinds of attacks are identified daily, and some remain undiscovered. In this work, we focus on Botnet based DDoS attacks that affect the application layer, especially the Web server [28]. The type of DDoS attack depends on the vulnerability of exploitation. The first type of attack is characterized by the consumption of the resources of the host. The victim can generally be a Web server or a proxy connected to the Internet. When the traffic load is high, packets are sent out to inform senders, who can either be legitimate users or attack sources, to reduce their sending rates. Legitimate users respond by decreasing their sending rates, whereas attack sources maintain or even increase their sending rates. Consequently, resources of the host, such as the CPU or memory capacity, become depleted, and the host is hindered from servicing legitimate traffic. The second type of attack involves the consumption of network bandwidth. If malicious traffic in the network dominates the communication links, traffic from legitimate sources is obstructed. In effect, bandwidth DDoS attacks are more disruptive than attacks resulting in resource consumption [29]. Detail discussion of these attacks is given below:

## 5.1 Net DDoS-based Bandwidth Attacks

Net DDoS-based bandwidth attacks are normally introduced effectively from a single attack source that takes advantage of specific IP weaknesses. Examples of such attacks are SYN and ICMP flood attacks.

### *5.1.1 SYN Flood Attacks*
A SYN flood attack utilizes a vulnerability of the TCP three-way handshake, such that a server must contain a large data structure for incoming SYN packets regardless of authenticity. During SYN flood attacks, SYN packets are sent by the attacker with unknown or non-existent source IP addresses. The three-way handshake occurs when the server stores the request information from the client into the memory stack and then waits for client confirmation. Given that the source IP addresses in SYN flood attacks are unknown or non-existent, confirmation packets for the requests created by the SYN flood attack are not received. Each half-open connection accumulates in the memory stack until it times out. Hence, the memory stack becomes full. Consequently, no requests can be processed, and the services of the system are disabled. Thus, SYN flood attacks are considered one of the most powerful flooding methods [30].

### *5.1.2 ICMP Flood Attacks*
ICMP is based on the IP protocol that can diagnose the status of the network. An ICMP flood attack is a bandwidth attack that uses ICMP packets that can be directed to an individual machine or to an entire network. When a packet is sent from a machine to an IP broadcast address in the local network, all machines in the network receive the packet. When a packet is sent from a machine to the IP broadcast address outside the local network, the packet is delivered to all machines in the target network. Other types of ICMP flood attack are the SMURF and the Ping-of-Death attacks [31].

## 5.2 App-DDoS Attacks

Attack power can be amplified by forcing the target to execute expensive operations. These attacks can consume all available corporate bandwidth and fill the pipes with illegitimate traffic. Routing protocols can also be affected and services are disrupted by either resetting the routing protocols or offering data that harm server operation [29].

### *5.2.1 HTTP Flood Attacks*

An attack that bombards Web servers with HTTP requests is called an HTTP flood attack. According to [32], HTTP flood attacks are common in most Botnet software programs. To send an HTTP request, a valid TCP connection that requires a genuine IP address has to be established. Attackers send an HTTP request through the IP address of a bot and then formulate the HTTP requests in different ways to maximize the attack power or to avoid detection. An attacker, for example, can manipulate the Botnet to send HTTP requests to download a large file from the target. The file is then read by the target from the hard disk, stored in the memory, and finally loaded into the packets, which are sent back to the Botnet. Hence, a simple HTTP request can significantly consume resources in the CPU, memory, input/output devices, and outbound Internet link.

However, the behavior of HTTP requests from the abovementioned example can be obvious. Repetitive requests for a large file can be detected and can then be blocked. Attackers mimic legitimate traffic by instructing the Botnet to send an HTTP request to the target Web site, analyze the replies, and then recursively follow the links. The HTTP requests from the attacker consequently become very similar to normal Web traffic, thus explaining the extreme difficulty in filtering this type of HTTP flood.

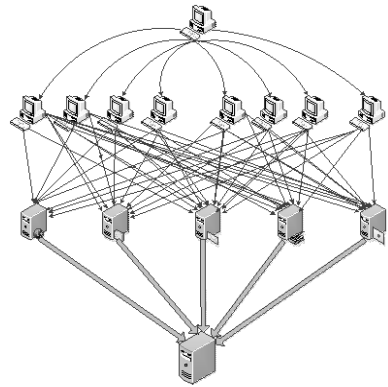

**Fig. 4: Distributed reflector denial of service (DRDoS) Attack**

### *5.2.2 Session Initiation Protocol (SIP) Flood Attacks*
The SIP is a widely supported standard for call set-up in Voice-over IP (VoIP). SIP proxy servers generally require public Internet access to accomplish the standard in accepting call set-up requests from any VoIP client. For scalability, SIP is typically implemented with UDP to become stateless. The attacker can flood the SIP proxy in one attack using SIP INVITE packets that pose as genuine source IP addresses. To avoid counter-hacking mechanisms, attackers can also launch the flood from a Botnet through a legitimate source IP address. Two victim categories emerge in this attack scenario. The first type comprises the SIP proxy servers with depleted server resources as a result of the processing of SIP INVITE packets, while their network capacity is consumed by the SIP INVITE flood. The SIP proxy server subsequently becomes incapable of providing VoIP service. The second type of victim is the call receiver, who becomes overwhelmed by fake VoIP calls and encounters difficulty in reaching legitimate callers [33].

### *5.2.3 Distributed Reflector Attacks*
Attackers should necessarily hide the true sources of their resulting attack traffic. Figure 4 illustrates the distributed reflector denial of service (DRDoS) attack, which hides attack traffic sources using third parties, such as routers or Web servers, during the relay of the attack traffic to the victim. These third parties are called reflectors. Any machine that responds to an incoming packet is a potential reflector. A DRDoS attack has three stages. In the second stage, after the attacker has gained control of "zombies," these "zombies" are instructed to send attack traffic information to the victims through the third parties, with the victim's IP address as the source IP address. In the third stage, the third parties send the reply traffic to the victim. This stage constitutes the DDoS attack. This type of attack had shut down a security research Web site (i.e. www.grc.com), in January 2002. DRDoS has been considered a potent and increasingly prevalent Internet attack. Unlike a traditional DDoS attack, the traffic from a DRDoS attack is further dispersed through third parties, resulting in the increased distribution of the attack traffic and increasing the difficulty in the identification of the attack. Moreover, the source IP addresses of the attack traffic point to innocent third parties, thus complicating the process of tracing the attack traffic source. Finally, as observed by [34] and [35], DRDoS attacks can amplify the attack traffic, thereby making the attack even more potent. In the succeeding section, an actual example demonstrates the serious threat posed by DRDoS attacks.

### *5.2.4 Domain Name System (DNS) Amplification Attacks*

An example of effective reflector attack is the DNS amplification attack shown in Figure 5. DNS provides a distributed infrastructure for the storage and association of different resource records (RR) with Internet domain names. DNS translates domain names into IP addresses. A recursive DNS server usually accepts a query and then resolves a given domain name for the requester. A recursive name server often contacts other authoritative name servers when necessary and subsequently returns the query response to the requester [36]. DNS query responses have disproportional sizes that normally comprise the original query and the answer. The query response packet is always larger than the query packet. Moreover, a query response can contain multiple RR, and some RR types can be very large.

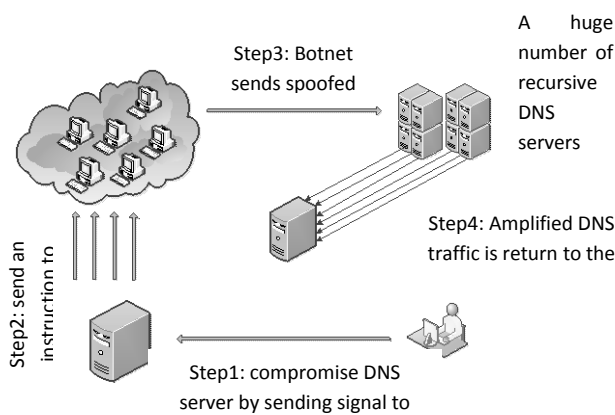

**Fig. 5: DNS Attack**

## 5.3 Trends that surprise in application-layer DDoS attacks

### *5.3.1 Bypass One Layer of Security*
In most cases, the applications that attackers are trying to exploit or target are well-known and must be "allowed" through perimeter security devices such as firewalls or IPS devices. For example, by default, firewalls allow HTTP or DNS traffic. IPS devices are not much different as they enforce security policy by inspecting packets for signatures of known threats. DDoS attacks take advantage of the fact that firewalls and IPS devices will pass legitimate traffic—thus eliminating one layer of security for the attacker.

### *5.3.2 Follow the Money*
Attackers see a major opportunity for extortion when applications are supporting high revenue-generating services. For example, an online gaming company is far more likely to pay an attacker to stop a DDoS attack that is costing millions per day in revenue than is an owner of a nonprofit Web site.

### *5.3.3 More Bang for the Buck*

Some attacks cause significantly more collateral damage than others. For example, a DNS attack that targets a single DNS service provider impacts not only that provider but all of its customers as well.

Organizations are beginning to realize that the power to rapidly stop application-layer DDoS attacks that target Internet-facing services is imperative for business continuity and success.

What makes this sort of attack different than a network or transport layer attack is that there is no way for upstream networking equipment to easily detect and filter out the attacker, since at the packet level, the traffic appears to be normal application traffic. System administrators and application developers of potential targets must instead take measures to build DDoS protection into their network and application design. Techniques such as caching and load balancing can increase the applications ability to absorb a flood of requests without becoming offline.

## 6. BOTNETS BASED DDOS ATTACK INCIDENTS

A DDoS attack is a major Internet threat as it can create a huge volume of unwanted traffic. DDoS attacks can prevent access to a particular resource, such as a Web site [37]. The first reported large-scale DDoS attack occurred in August 1999 against a University [38]. The attack had shut down the victim's network for over two days. In 7 February 2000, a number of Websites went offline for several hours after an attack [38]. In some cases, DDoS attacks can produce approximately 1 Gbit/s of attack traffic against a single victim [39]. In February 2001, over 12,000 attacks were registered against more than 5,000 distinct victims over a three-week period [40]. The Coordination Center of the Computer Emergency Response Team was also attacked in May 2001, making the availability of their Website intermittent for more than two days.

DDoS attacks usually continuously target DNS. In October 2002, all root name servers underwent an exceptionally intensive DoS attack [41] with some non-received DNS requests to an outsourced DNS service in Akamai, which were meant to enhance service performance. In 2004, UK online bookmaking, betting, and gambling sites were overwhelmed by DoS attacks launched by unidentified attackers. The Internet-based business service of Al Jazeera, a provider of Arabic-language news services, was similarly attacked in January 2005. The text-to-speech translation application in the Sun Microsystem's Grid computing system was disabled during its opening day by a DoS attack in March 2006.

In [39], the presence of roughly 2,000 to 3,000 active DoS attacks per week was described using an updated backscatter analysis. The attack record over a three-year period revealed 68,700 attacks on over 34,700 distinct Internet hosts from more than 5,300 organizations. Some DNS requests failed to reach a root name server because of the congestion caused by the DoS attack. In [41], another major DoS attack occurred on 15 June 2004 against name servers in the Akamai Content Distribution Network. This attack blocked almost all access to such sites as Apple computer, Google, Microsoft, and Yahoo for more than two hours. These companies supposedly outsourced their DNS service to Akamai for improved performance.

## 6.1 Recent Botnet based DDoS Incidents

DDoS attacks occur almost daily. Even well-known websites, such as Twitter, Facebook, Google, and other popular search engines, cannot escape these attacks that affect countless users. An eye-opener case was the DDoS incident that targeted the White House, FBI, DOJ [42], the Recording Business Association of America, Universal Music Websites, and the Hong Kong Stock Exchange [43]. A total of 80 computers were compromised by the Botnet and up to 250,000 were infected with malware during the attack. The attack traffic consumed 45 gigabytes per second according to the 7th Annual Report from the Arbor Company 2011 [5]. The outage lasted for seven days, the longest in 2010. In 2011, the longest attack ever recorded target a travel company, lasting for 80 days, 19 hours, 13 minutes, and 5 seconds. The average duration of DDoS attacks is 9 hours and 29 minutes. The observed DDoS incidents from 2011 to the first quarter of 2012 are shown in Table 1.

## 6.2 Financial Losses by Botnet based DDoS Attack

Large-scale attacks cause substantial financial damage to companies relying on the Internet for their daily business. Direct (e.g., revenue loss during the attack) and indirect (e.g., customer loss attributed to degraded reputation) damages are also experienced. E-commerce and stock exchange sites spend millions of dollars to recover from these attacks, whereas other companies allocate a huge amount of money to defend themselves from possible hackers. As indicated by the survey of VeriSign respondents, expenditures reach up to $2.5 million [57]. Table 2 shows the loss of revenue attributed to service disruption among large companies in the world.

## 7. CONCLUSIONS

A Botnet-based DDoS attack is undoubtedly a serious Internet problem that challenges the growth rate and the public acceptance of online government and business sites. In this paper, a clear view of the Botnet based DDoS attack on the application layer, especially on the Web server, is presented. Incidents around the world and revenue losses of famous companies and government Web sites are also described, indicating that extreme care should be taken and a further study should be conducted to assess the size of the problem and then derive an optimal solution.

## 8. ACKNOWLEDGMENTS

The authors gratefully acknowledge the financial support of the National Advanced IPv6 Centre (NAV6), Universiti Sains Malaysia, for partial work reported in the paper.

**Table 1. Botnet-Based DDoS attack incidents 2011-2012**

| The Target | Date of Attack | Details |
|---|---|---|
| Tunisian Government Web sites | 3 January 2011 | Web site outage that included the president, prime minister, ministry of industry, ministry of foreign affairs, and stock exchange [44] |
| FINE GAEL's News Web site www.finegael2011.com | 9 January 2011 | One-night content outage by an anonymous attacker using the LOIC tool [45] |
| Egyptian government Web sites | 25 January 2011 | Site went offline from the beginning of the revolution until the president stepped down [46] |
| HB Gary Federal | 5–6 February 2011 | Hacked by dumping 68,000 e-mails from the system [47] |
| Operation Ouraborus | 16 February 2011 | Threats from an anonymous attacker who hacked the site and caused irreversible damage [48] |
| NEW YORK (CNN Money) | 3 March 2011 | The huge attack hit the company's data centers with tens of millions of packets per second [49] |
| Operation Empire State Rebellion | 14 March 2011 | Threat from an anonymous attacker affecting the Bank of America [50] |
| Operation Sony | April 2011 | Outage of the Play Station Network [51] |
| Spanish Police | 12 Jun 2011 | DDoS attack lasted for approximately one hour [52] |
| Operation Malaysia Malaysia.gov.my | 15 Jun 2011 | Outage of 91 Web sites of the Malaysian Government that started 7:30 pm GMT [53] |
| Operation Orlando | 16 Jun 2011 | Orlando government Web sites went offline daily because of the LOIC tool [54] |
| Visa Card, Master Card, Wikileaks and www.paypal.com | 27 July 2011 | Payment processing from Wikileaks through PayPal were continuously denied [55] |
| Hong Kong stock exchange | 15 August 2011 | Hundreds of companies were affected with a single target [56] |
| Justice.gov, MPAA.org, White House, the FBI, BMI.com, Copyright.com, Viacom, Anti-piracy.be/nl, Vivendi.fr, Hadopi.fr, and ChrisDodd.com, | 19 Jan 2012 | The largest attack for 2012 from an anonymous attacker who shut down all the affected sites for 10 minutes [22] |

**Table 2. Financial Losses by Botnet based DDoS Attack [58]**

| Action | Company A | Company B | Company C | Company D |
|---|---|---|---|---|
| Loss of Revenue per hour | $19 million | $240,000 | $650,000 | $190,000 |
| Line Business | E-banking | E-banking | E-Commerce | E-Commerce |
| Average Loss in an attack | $38-114 million | $480,000-1,440,000 | $1,300,000-3,900,000 | $380,000-1,140,000 |